\documentclass[a4paper,journal]{IEEEtran}

\IEEEoverridecommandlockouts
\usepackage{graphicx}
\usepackage{amssymb}
\usepackage{amsmath}
\usepackage{algcompatible}
\usepackage{algorithm}
\usepackage{algpseudocode}
\usepackage{algpseudocodex}
\usepackage{array}
\usepackage{enumitem}
\usepackage{textcomp}
\usepackage{physics}
\usepackage{stfloats}
\usepackage{url}
\usepackage[draft]{hyperref}
\usepackage{verbatim}
\usepackage{booktabs}
\usepackage{tabularx}
\usepackage{cite}
\usepackage[short, c3]{optidef}
\usepackage[mathlines]{lineno}
\usepackage{color}
\usepackage{soul}
\usepackage{tabulary}
\usepackage{multirow}
\usepackage{pbox}
\usepackage{multicol}
\usepackage{lipsum}
\usepackage{amsthm}
\usepackage{relsize}
\usepackage{epstopdf}
\usepackage{mathtools}
\usepackage{calc}
\usepackage{makecell}
\usepackage{pgfplots}\pgfplotsset{compat=1.18}
\usepackage[caption=false]{subfig}

\usepackage{algpseudocode}
\usepackage{mathtools}
\usepackage{pdfpages}

\clearpage{\pagestyle{empty}\cleardoublepage}
\newcommand{\beq}{\begin{equation}}
\newcommand{\eeq}{\end{equation}}
\newcommand{\bitm}{\begin{itemize}}
\newcommand{\ba}{\begin{array}}
\newcommand{\ea}{\end{array}}
\newcommand{\eitm}{\end{itemize}}
\newcommand{\beqn}{\begin{eqnarray}}
\newcommand{\eeqn}{\end{eqnarray}}
\newcommand{\beqno}{\begin{eqnarray*}}
\newcommand{\eeqno}{\end{eqnarray*}}
\newcommand{\bma}{\begin{displaymath}}
\newcommand{\ema}{\end{displaymath}}
\newcommand{\bnu}{\begin{enumerate}}
\newcommand{\enu}{\end{enumerate}}
\newcommand{\bce}{\begin{center}}
\newcommand{\ece}{\end{center}}
\newcommand{\btb}{\begin{tabular}}
\newcommand{\etb}{\end{tabular}}

\usepackage{graphicx}
\usepackage{textcomp}
\usepackage{xcolor}
\def\BibTeX{{\rm B\kern-.05em{\sc i\kern-.025em b}\kern-.08em
    T\kern-.1667em\lower.7ex\hbox{E}\kern-.125emX}}

\begin{document}

\addtolength{\textwidth}{-0.4cm}
\addtolength{\oddsidemargin}{0.2cm}
\addtolength{\evensidemargin}{0.2cm}

\title{\huge On the Investigation of Antenna Beam Pattern Models for Fast Beam Placement in LEO Satellite Networks
}

\author{
Ninh Thi-Thanh Tam$^{*, \xi}$,
Nguyen Viet Anh$^{\xi}$,
Nguyen Minh Quan$^{\xi}$,
Trinh Van Chien$^{\xi}$%
\\[2pt]
{\fontsize{9}{9}\selectfont
$^{*}$Faculty of Information and Communication Technology, National Academy of Education Management, Vietnam}%
\\
{\fontsize{9}{9}\selectfont
$^{\xi}$School of Information and Communications Technology, Hanoi University of Science and Technology, Vietnam}%

{\fontsize{9}{9}\selectfont Corresponding author: Trinh Van Chien}

}
\maketitle
\begin{abstract}
In this paper, we investigate the impact of antenna beam pattern models on the beam placement problem in Low Earth Orbit (LEO) satellite networks. Unlike conventional beam placement studies that often rely on a fixed antenna pattern or a predefined half-power beamwidth, this work compares four representative gain models, namely the Bessel $J_1$ model, the combined Bessel $J_1+J_3$ model, the Gaussian model, and a 3GPP-TR 37.842 model. For each antenna pattern, the corresponding half-power angle is obtained and used to construct a user compatibility graph. The beam placement problem is then transformed into a Minimum Clique Cover problem and solved using a graph coloring-based algorithm. Numerical results compare the considered antenna models in terms of the number of active beams, load balancing, statistical channel gain-to-noise ratio (SCGNR), and signal-to-interference-plus-noise ratio (SINR). We analyze the impact of increasing the number of active beams on SINR-based spectral efficiency to explain the importance of carefully controlling the number of active beams. The results show that antenna beam pattern models affect both beam resource usage and link quality, highlighting the need for antenna-pattern-aware beam placement in LEO satellite networks.
\end{abstract}


\thispagestyle{empty}

\section{Introduction}
\label{sec:introduction}

Low Earth Orbit (LEO) satellite networks have recently attracted significant attention as a key enabling technology for future global broadband access, massive Internet-of-Things (IoT) connectivity, emergency communications, and sixth-generation (6G) non-terrestrial networks \cite{hui2025review}. Com with geostationary satellite systems, LEO satellites operate at much lower altitudes, which leads to reduced propagation delay, lower free-space path loss, and improved link quality \cite{gao2025performance}. These advantages make LEO satellite systems particularly suitable for providing high-capacity and low-latency services to users located in remote, rural, maritime, and disaster-affected areas where terrestrial infrastructure is unavailable or unreliable. To efficiently serve a large number of geographically distributed ground users, modern LEO satellites usually employ multibeam antenna architectures. By generating multiple spot beams, the satellite can spatially divide its coverage region into smaller service areas and reuse radio resources across different beams \cite{thi2025clustering}. However, the design and placement of these beams are challenging since the number of available beams, transmit power, bandwidth, and onboard processing capability are limited \cite{cardakli2025challenges}. On the other hand, using too few beams may degrade link quality, increase user load imbalance, and reduce the ability of the system to serve spatially distributed users \cite{cai2025energy}. Therefore, fast and resource-efficient beam placement is a fundamental problem in LEO multibeam satellite networks.

The beam placement problem determines user grouping and beam center locations. In practice, different antenna beam pattern models may have different main-lobe widths, roll-off characteristics, and sidelobe behaviors, leading to different effective beam coverage regions. Since the received antenna gain depends on the angular radiation pattern, the antenna model directly affects the half-power angle of a beam \cite{li2026electrically}. This angle can be used to determine user compatibility, i.e., whether two users can be potentially served by the same beam while maintaining acceptable antenna gain. Consequently, different antenna patterns may produce different compatibility graphs, user groupings, and numbers of active beams. Furthermore, the number of active beams strongly affects system performance. While more beams can reduce the distance between users and their serving beam centers, they also divide the limited satellite transmit power and increase inter-beam interference under full frequency reuse. Therefore, activating more beams does not necessarily improve SINR or spectral efficiency, revealing a trade-off among coverage granularity, power sharing, interference, and resource utilization.

Existing studies on multibeam satellite networks have addressed a wide range of resource management problems, including user association, beam hopping, power allocation, bandwidth allocation, interference coordination, and throughput maximization. In the context of LEO satellite networks, beam placement is particularly challenging due to the rapid movement of satellites, time-varying user visibility, limited onboard resources, and the requirement for fast beam reconfiguration. Several recent works have investigated beam placement from different perspectives. For example, some studies mainly focus on the spatial distribution of ground users and design beam layouts according to user density or geographical clustering \cite{chan2025near}. These methods are effective in adapting the beam layout to non-uniform traffic demand, but the beam coverage is usually modeled by a predefined beam radius or a fixed half-power beamwidth. Other works study classification- or learning-based approaches to identify beam placement patterns, user grouping structures, or service regions in LEO satellite systems \cite{eftetahi2026fair}. These approaches improve the adaptability of beam management, but they still commonly rely on simplified antenna coverage assumptions. In addition, a number of studies consider beam hopping, power control, and interference coordination after the beam layout has been determined \cite{kim2025multi}. Although these methods improve resource utilization and link performance, the antenna beam pattern used to construct the beam coverage region is often treated as fixed or is not explicitly analyzed.

In practical multibeam satellite systems, the antenna beam pattern strongly affects the effective coverage region of each beam. However, existing fast beam placement studies in LEO satellite networks often assume a fixed beamwidth and have not systematically compared different antenna models. To address this gap, this paper provides an antenna-pattern-aware analysis of fast beam placement. The main contributions are threefold: (i) we investigate the impact of four representative antenna beam pattern models, namely Bessel $J_1$, Bessel $J_1+J_3$, Gaussian, and 3GPP-TR 37.842 \cite{3GPP_TR37842}, on fast beam placement in LEO satellite networks; (ii) we develop an antenna-pattern-aware implementation of the graph-based beam placement framework, where the half-power angle of each antenna model is used to construct the user compatibility graph and the resulting problem is solved as a Minimum Clique Cover problem via graph coloring; and (iii) we compare the considered models in terms of the number of active beams, load balancing, SCGNR, SINR, and SINR-based spectral efficiency under different active-beam configurations.
\vspace{-3pt}
\section{System Model and Antenna Gain Models}
\label{sec:systemmodel}
\vspace{-3pt}
\subsection{System Model}
\label{sec:sys_model}

\begin{figure}[t]
    \label{fig:model}
    \centering
    \includegraphics[trim={0cm 0cm 0.0cm 0.0cm},clip,width=0.4\textwidth]{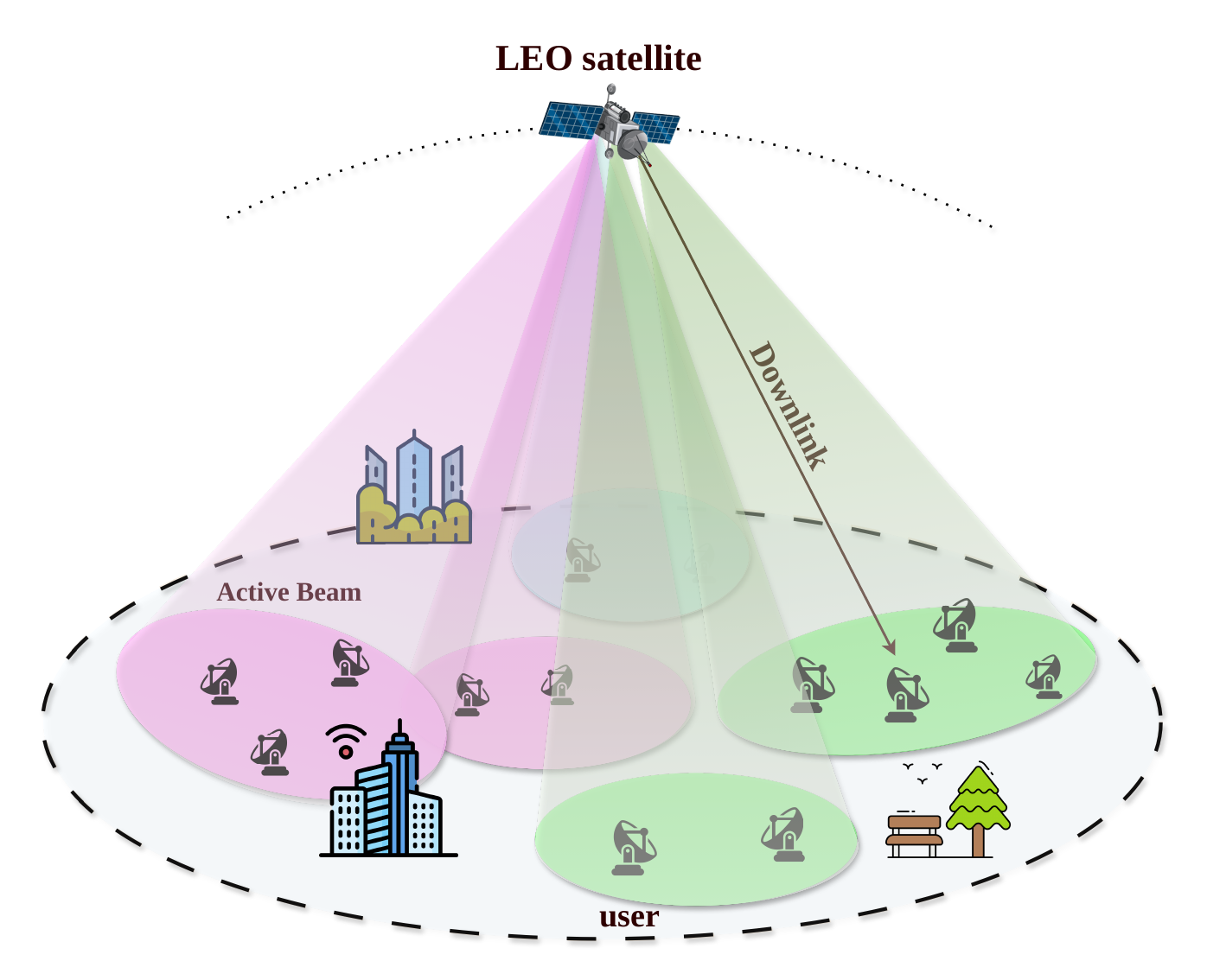}
    \caption{\centering{Multibeam coverage and user distribution in the LEO system}
    }
\end{figure}

We consider a downlink LEO satellite communication system in which a single satellite serves a set of ground users through multiple spot beams as shown in Fig.~\ref{fig:model}. Let $\mathcal{T}=\{1,2,\ldots, K\}$ denote the set of ground users randomly distributed within the satellite coverage area. The satellite is equipped with a maximum of $N$ available spot beams, of which only a subset is activated to provide communication services. Let $B$ represent the number of active beams, and define the corresponding active-beam set as $\mathcal{B}=\{1,2,\ldots,B\}$, where $B\leq N$.

Each user is associated with exactly one active beam, while each active beam can simultaneously serve multiple users. Let $\mathcal{T}_b\subseteq\mathcal{T}$ denote the set of users served by beam $b$. Accordingly, the user association must satisfy the following constraints:
\begin{equation}
\bigcup_{b\in\mathcal{B}}\mathcal{T}_b=\mathcal{T},\quad
    \mathcal{T}_b\neq\emptyset,\quad
    \mathcal{T}_b\cap\mathcal{T}_{b'}=\emptyset, |\mathcal{T}_b| \geq 0, \ \forall b\neq b'.
    \label{eq:user_partition}
\end{equation}
The objective of the fast beam placement problem is to determine a resource-efficient beam layout, guaranteeing full user coverage while minimizing the number of active beams.

Let $\mathbf{q}_k$ denote the geographical location of user $k$ and let $\mathbf{c}_b$ represent the ground projection of the center of beam $b$. Furthermore, let $D_k$ and $D_b$ denote the slant-range distances from the satellite to user $k$ and beam center $b$, respectively. Define $d_{b,k}$ as the geographical distance between $\mathbf{q}_k$ and $\mathbf{c}_b$, the off-boresight angle $\alpha_{b,k}$ measured between user $k$ and the center of beam $b$ from the satellite's perspective, can be determined using the Spherical Law of Cosines as
\begin{equation}
    \alpha_{b,k} = \arccos\left(\frac{D_k^2 + D_b^2 - d_{b,k}^2}{2D_k D_b}\right).
\end{equation}
User $k$ is regarded as being covered by beam $b$ if the antenna gain corresponding to its angular position is no less than one-half of the maximum beam gain. Specifically, let $g_{b,k}$ denote the antenna gain between beam $b$ and user $k$, and let $G_{\max}$ denote the maximum antenna gain at the beam boresight. The half-power beam coverage condition is therefore given by $g_{b,k}\geq G_{\max}/2$. In this work, the antenna gain as
\begin{equation}
    g_{b,k}^{(m)}=G_{\max}F_m(\alpha_{b,k}),
    \label{eq:gain_general}
\end{equation}
where $F_m(\cdot)$ represents the normalized antenna radiation pattern associated with antenna model $m$. To comprehensively evaluate the proposed beam placement framework under different antenna characteristics, four representative antenna gain models are considered: the Bessel $J_1$ model, the combined Bessel $J_1+J_3$ model, the Gaussian model, and the 3GPP TR 37.842 model.
The statistical channel gain between beam $b$ and user $k$ is expressed as
\begin{equation}
    h_{b,k}^{(m)}
    =
    \frac{g_{b,k}^{(m)}\bar{g}_{k}}
    {L_{\mathrm{fs},k}L_{\mathrm{atm}}},
    \label{eq:scg}
\end{equation}
where $\bar{g}_{k}$ denotes the receive antenna gain, $L_{\mathrm{fs},k}$ represents the free-space path loss, and $L_{\mathrm{atm}}$ accounts for the atmospheric attenuation. The free-space path loss is expressed as
\begin{equation}
    L_{\mathrm{fs},k} = \frac{16\pi^2 D_k^2}{\lambda^2},
\end{equation}
where $\lambda$ is the carrier wavelength. The SCGNR of user $k$ served by beam $b_k$ under antenna model $m$ is defined as
\begin{equation}
    \xi_k^{(m)}
    =
    \frac{h_{b_k,k}^{(m)}}{\sigma^2},
    \label{eq:scgnr}
\end{equation}
where $\sigma^2$ denotes the additive white Gaussian noise (AWGN) power. When all active beams reuse the same frequency band, inter-beam interference is taken into account. Assuming equal transmit power allocation among the active beams, the transmit power assigned to beam $b$ is $P_b=P_{\mathrm{sat}}/{B}$,
where $P_{\mathrm{sat}}$ is the total satellite transmit power. The SINR and spectral efficiency of user $k$ served by beam $b_k$ is
\begin{align}
    \gamma_k^{(m)}
    &=
    \frac{P_{b_k}h_{b_k,k}^{(m)}}
    {\sigma^2+\sum_{j\in\mathcal{B},j\neq b_k}P_jh_{j,k}^{(m)}},
    \label{eq:sinr}
    \\
    \mathrm{SE}_k^{(m)}&=\log_2\left(1+\gamma_k^{(m)}\right).
    \label{eq:spectral_efficiency}
\end{align}

\subsection{Antenna Beam Pattern Models}
\label{sec:channel_assumption}

This subsection introduces the four antenna beam pattern models considered in this work. Each model is characterized by a distinct normalized antenna gain function, $F_m(\alpha)$, which determines the half-power beamwidth and consequently influences the beam placement strategy.

\subsubsection{Bessel \texorpdfstring{$J_1$}{J1} Model}

The first antenna model is derived from the radiation pattern of a circular aperture antenna. Its normalized gain function is given by \cite{van2024fast}
\begin{equation}
    F_{J_1}(\alpha)
    =
    4\left|
    \frac{J_1(u)}{u}
    \right|^2,
    \label{eq:j1_model}
\end{equation}
where $J_1(\cdot)$ is the first-order Bessel function of the first kind, and $u=\frac{2\pi r}{\lambda}\sin(\alpha).$
Here, $r$ is the radius of the antenna aperture. To ensure numerical stability, the value of $u$ is lower-bounded by a small positive constant when $\alpha$ approaches zero.

\subsubsection{Bessel \texorpdfstring{$J_1+J_3$}{J1+J3} Model}

The second antenna model uses a combined Bessel pattern involving both the first- and third-order Bessel functions. The corresponding normalized gain function is expressed as \cite{choi2023joint}
\begin{equation}
    F_{J_1+J_3}(\alpha)
    =
    \left[
    \frac{J_1(u)}{2u}
    +
    36\frac{J_3(u)}{u^3}
    \right]^2,
    \label{eq:j1j3_model}
\end{equation}
where $J_3(\cdot)$ is the third-order Bessel function of the first kind. Compared with the conventional $J_1$ model, this model produces a different main-lobe profile, resulting in a different effective half-power beamwidth.

\subsubsection{Gaussian Model}

The Gaussian model provides a smooth and rapidly decaying main-lobe approximation. Its normalized gain is expressed as \cite{angervuori2025meta} ${F_{\mathrm{G}}(\alpha)
    =
    2^{-u^2},}$
when $u$ corresponds to the half-power angle, the normalized gain satisfies $F_{\mathrm{G}}(\alpha) = 1/2$.

\subsubsection{3GPP-TR 37.842 Model}

The fourth model is a simplified 3GPP-TR 37.842 antenna pattern. In the dB domain, the attenuation is commonly represented by a quadratic function of the angular deviation. In linear scale, the normalized gain can be written as \cite{recommendation20172101} ${F_{\mathrm{3GPP}}(\alpha) = 10^{-1.2 u^2},}$. With this definition, at the half-power reference angle, the gain evaluates to $F_{\mathrm{3GPP}}(\alpha) = 10^{-0.3} \approx \frac{1}{2}.$

\subsection{Half-Power Angle Extraction}
\label{sec:half_power_extraction}

For each antenna gain model $m$, the half-power angle is defined as the angular offset at which the normalized gain decreases to one half of its maximum value. Mathematically, it is determined by solving $F_m(u_{\mathrm{HP},m})=1/2$, where the corresponding half-power angle is obtained as $\theta_{\mathrm{HP},m}=\arcsin\left(\frac{u_{\mathrm{HP},m}\lambda}{2\pi r}\right)$. The corresponding full half-power beamwidth is then $\alpha_{\max,m}=2\theta_{\mathrm{HP},m}$. The extracted half-power angle is used to determine whether two users can be grouped into the same beam. Specifically, for two users $i$ and $j$, let $\alpha_{i,j}$ denote their angular separation observed from the satellite. Under the antenna model $m$, the pairwise compatibility condition is $\alpha_{i,j}\leq \theta_{\mathrm{HP},m}$. This condition indicates that users $i$ and $j$ can be potentially covered by the same beam while remaining within the half-power region. Accordingly, different antenna beam pattern models can produce different half-power angles, resulting in different user compatibility graphs and, consequently, different beam placement outcomes. This allows us to investigate how antenna beam pattern assumptions affect the number of active beams, load balancing, SCGNR, and SINR performance.

\section{Graph-Based Fast Beam Placement}
\label{sec:problem_Formulation}

\subsection{User Compatibility Graph}

For each antenna beam pattern model $m$, the half-power angle 
$\theta_{\mathrm{HP},m}$ is obtained from $F_m(\theta_{\mathrm{HP},m})=1/2$. This angle determines whether two users can be grouped into the same beam. 
Let $\alpha_{i,j}$ denote the angular separation between users $i$ and $j$ observed from the satellite. An undirected user compatibility graph $\mathcal{G}_m=(\mathcal{V},\mathcal{E}_m)$ is constructed, where each vertex represents one user, i.e., $\mathcal{V}=\mathcal{T}$. An edge exists between two users if their angular separation is no larger than the half-power angle. Accordingly, the adjacency matrix $\mathbf{A}^{(m)}$ is defined as
\begin{equation}
    a_{i,j}^{(m)}
    =
    \begin{cases}
        1, & \alpha_{i,j}\leq \theta_{\mathrm{HP},m}, \; i\neq j,\\
        0, & \text{otherwise}.
    \end{cases}
    \label{eq:adjacency_matrix}
\end{equation}

\subsection{Minimum Clique Cover Formulation}

A group of users can be served by a single beam if every pair of users inside the group is mutually compatible. In graph-theoretic terms, this means that the corresponding vertices must form a clique in $\mathcal{G}_m$. Let $\mathcal{T}_b$ denote the set of users assigned to beam $b$. Then, $\mathcal{T}_b$ is feasible if $a_{i,j}^{(m)}=1, \forall i,j\in\mathcal{T}_b,\ i\neq j.$
Thus, each feasible user group corresponds to a clique in the compatibility graph. The fast beam placement problem aims to partition all users into the minimum number of feasible user groups. This is equivalent to covering all vertices of $\mathcal{G}_m$ by the minimum number of disjoint cliques. The problem can be written as
\begin{align}
    \min_{\{\mathcal{T}_b\}_{b=1}^{B}} \quad & B \\
    \text{s.t.} \quad
    & \bigcup_{b=1}^{B}\mathcal{T}_b=\mathcal{T}, \\
    & \mathcal{T}_b\cap\mathcal{T}_{b'}=\emptyset,\quad b\neq b', \\
    & \mathcal{T}_b \text{ forms a clique in } \mathcal{G}_m,\quad \forall b.
\end{align}
This problem is known as the Minimum Clique Cover (MCC) problem. Since MCC is NP-hard, we solve it approximately by transforming it into a graph coloring problem on the complement graph. Let $\overline{\mathcal{G}}_m=(\mathcal{V},\overline{\mathcal{E}}_m)$ be the complement graph of $\mathcal{G}_m$. In $\overline{\mathcal{G}}_m$, two vertices are connected if the corresponding users are not compatible in $\mathcal{G}_m$. 
In this complement graph, two incompatible users must be assigned different colors. Therefore, a proper coloring of $\overline{\mathcal{G}}_m$ produces a clique cover of $\mathcal{G}_m$, where each color corresponds to one active beam. To solve the graph coloring problem, a greedy coloring algorithm is employed. The main idea is from \cite{van2024fast} and summarized in Algorithm~\ref{alg:graph_based_beam_placement}.



\vspace{0.2cm}
\noindent\textbf{Computational Complexity:} The complexity of Algorithm~\ref{alg:graph_based_beam_placement} is $\mathcal{O}(N^{3}K^{2})$ with $K$ users and $N$ maximum number of beams. Overall, we provide the performance guarantee for the solutions delivered by Algorithm~\ref{alg:graph_based_beam_placement}.

Since the MCC problem is NP-hard, we approximately solve it by
transforming the compatibility graph into its complement graph and
applying a greedy graph-coloring procedure. Each color corresponds
to one active beam. A subsequent refinement step redistributes
compatible users among beam groups to improve load balancing without
increasing the number of active beams. The computational complexity
follows the graph-based framework in \cite{van2024fast}.

\begin{algorithm}[t]
    \caption{Antenna-Pattern-Aware Graph-Based Beam Placement}
    \label{alg:graph_based_beam_placement}
    \begin{algorithmic}[1]
        \State \textbf{Input}: User set $\mathcal{T}$, angular separations $\{\alpha_{i,j}\}$, antenna model $m$, half-power angle $\theta_{\mathrm{HP},m}$.
        \State \textbf{Output}: Active beam set $\mathcal{B}$ and user grouping $\mathcal{S}$.

        \Statex \textit{Phase 1: Compatibility graph construction}
        \State Initialize adjacency matrix $\mathbf{A}^{(m)}=\mathbf{0}$.
        \ForAll{user pairs $(i,j)$, $i\neq j$}
            \If{$\alpha_{i,j}\leq \theta_{\mathrm{HP},m}$}
                \State Set $a_{i,j}^{(m)}=1$ and $a_{j,i}^{(m)}=1$.
            \EndIf
        \EndFor
        \State Construct the graph $\mathcal{G}_m=(\mathcal{V},\mathcal{E}_m)$, $\overline{\mathcal{G}}_m$.

        \Statex \textit{Phase 2: Greedy graph coloring}
        \State Sort all users in descending order of their degrees in $\overline{\mathcal{G}}_m$.
        \State Initialize the color of each user as unassigned.
        \ForAll{user $k$ in the sorted list}
            \State Assign to user $k$ the smallest color that is not used by its colored neighbors in $\overline{\mathcal{G}}_m$.
        \EndFor
        \State Group users with the same color into beam clusters.
        \State Let $\mathcal{S}=\{\mathcal{T}_1,\mathcal{T}_2,\ldots,\mathcal{T}_B\}$ be the obtained user groups.
        \State Set $\mathcal{B}=\{1,2,\ldots,B\}$.

        \Statex \textit{Phase 3: Load balancing refinement}
        \State $\mathrm{stop}=\mathrm{False}$.
        \While{not $\mathrm{stop}$}
            \State $\mathrm{count\_moves}=0$.
            \ForAll{pairs of user groups $(\mathcal{T}_b,\mathcal{T}_{b'})$}
                \If{$|\mathcal{T}_b|-|\mathcal{T}_{b'}|>1$}
                    \ForAll{user $k\in\mathcal{T}_b$}
                        \If{$a_{k,u}^{(m)}=1,\ \forall u\in\mathcal{T}_{b'}$}
                            \State Move user $k$ from $\mathcal{T}_b$ to $\mathcal{T}_{b'}$.
                            \State $\mathrm{count\_moves}=\mathrm{count\_moves}+1$.
                            \If{$|\mathcal{T}_b|-|\mathcal{T}_{b'}|\leq 1$}
                                \State \textbf{break}
                            \EndIf
                        \EndIf
                    \EndFor
                \EndIf
            \EndFor
            \If{$\mathrm{count\_moves}=0$}
                \State $\mathrm{stop}=\mathrm{True}$.
            \EndIf
        \EndWhile

        \State Compute the beam center of each group $\mathcal{T}_b$.
        \State \textbf{Return}: $\mathcal{B}$ and $\mathcal{S}$.
    \end{algorithmic}
\end{algorithm}

\section{Numerical Results}
\label{sec:numerical_results}


\subsection{Simulation Setup}

We consider a LEO satellite serving ground users randomly distributed within its coverage area, with the number of users set as $K\in\{100,250,500,1000\}$ and the total parameters are referred to in \cite{van2024fast}. For all considered antenna models, the physical aperture-to-wavelength ratio is unified as $r/\lambda=5$. For each user distribution and antenna pattern, the half-power angle is used to construct the user compatibility graph based on the angular separation between user pairs. The proposed graph-based beam placement method formulates the problem as a Minimum Clique Cover problem and solves it via graph coloring, where each color represents one active beam. The evaluated metrics include the average number of active beams, load balancing gap, SCGNR distribution, and SINR-based spectral efficiency. For spectral efficiency evaluation, equal power allocation is applied among active beams, and each user's SINR is computed by considering the serving beam, thermal noise, and inter-beam interference.

\subsection{Antenna Beam Pattern Comparison}

\begin{figure}[t]
    \centering
    \includegraphics[trim={1.2cm 8.7cm 0.0cm 9.0cm},clip,width=0.55\textwidth]{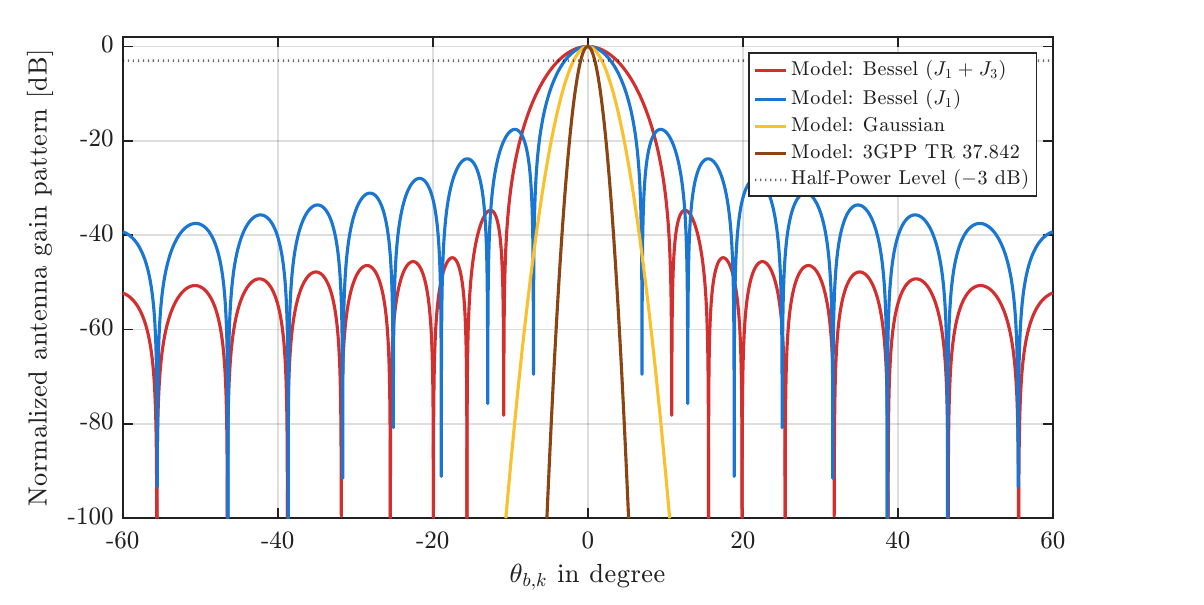}
    \caption{Normalized antenna gain patterns of the considered beam models.}
    \label{fig:beam_pattern_comparison}
\end{figure}

Fig.~\ref{fig:beam_pattern_comparison} compares the normalized antenna gain patterns of the four considered models. The Bessel-based models exhibit oscillatory sidelobe behavior, while the Gaussian and 3GPP-TR 37.842 models provide smoother main-lobe attenuation. The half-power level is marked at $-3$ dB. Different antenna patterns lead to different half-power angles, which directly affect the density of the user compatibility graph and the resulting beam placement. Specifically, Bessel-based radiation models yield a larger half-power angle $\theta_{\mathrm{HP},m}$, which relaxes the pairwise compatibility condition and expands the feasible domain for user clustering. Consequently, the resulting compatibility graphs under the Bessel-based models achieve a higher edge density compared to those obtained from the more conservative Gaussian and 3GPP-TR 37.842 counterparts.

\subsection{Impact on the Number of Active Beams}

\begin{figure}[t]
    \centering
    \includegraphics[trim={1.4cm 8.8cm 0.0cm 9.0cm},clip,width=0.55\textwidth]{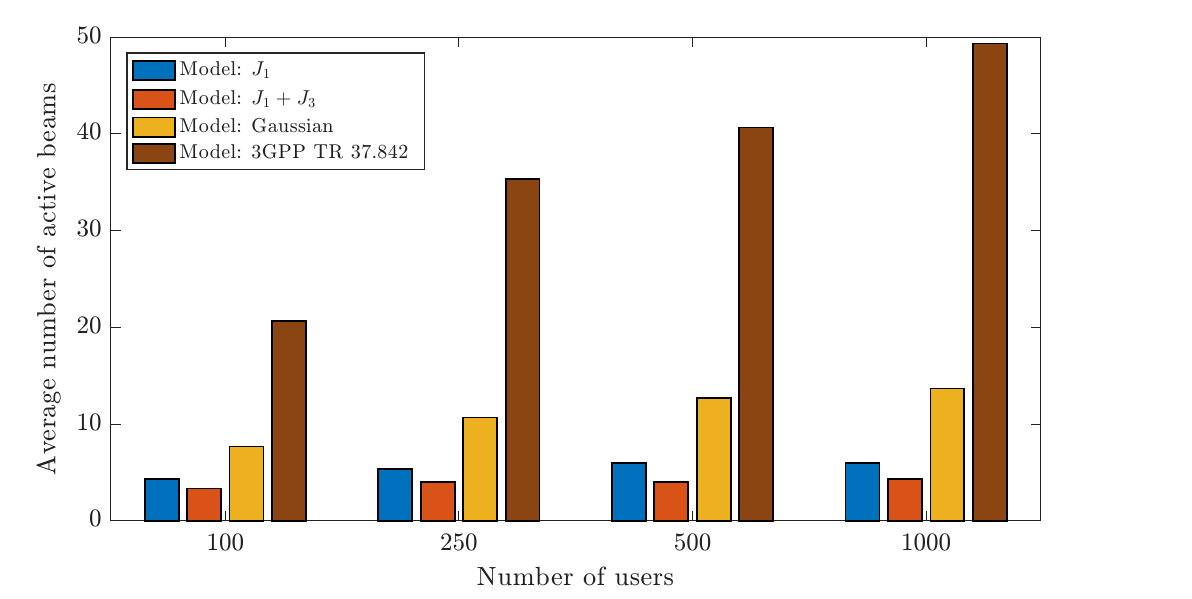}
    \caption{Average number of active beams versus the number of users.}
    \label{fig:nab_comparison}
\end{figure}

Fig.~\ref{fig:nab_comparison} shows the average number of active beams obtained by the graph-based beam placement algorithm. The results demonstrate that antenna beam patterns have a significant impact on the required number of active beams. In general, a model with a wider effective half-power region produces a denser compatibility graph, allowing more users to be grouped into the same beam. As a result, fewer active beams are required. Among the considered models, the Bessel-based models require fewer active beams, especially the combined $J_1+J_3$ model. The Gaussian model requires a moderate number of beams, while the 3GPP-TR 37.842 model leads to the largest number of active beams. This indicates that different beam attenuation characteristics can strongly affect the grouping capability of the fast beam placement algorithm.

\subsection{Impact on Load Balancing}

\begin{figure}[t]
    \centering
    \includegraphics[trim={1.7cm 8.8cm 0.0cm 9.0cm},clip,width=0.5525\textwidth]{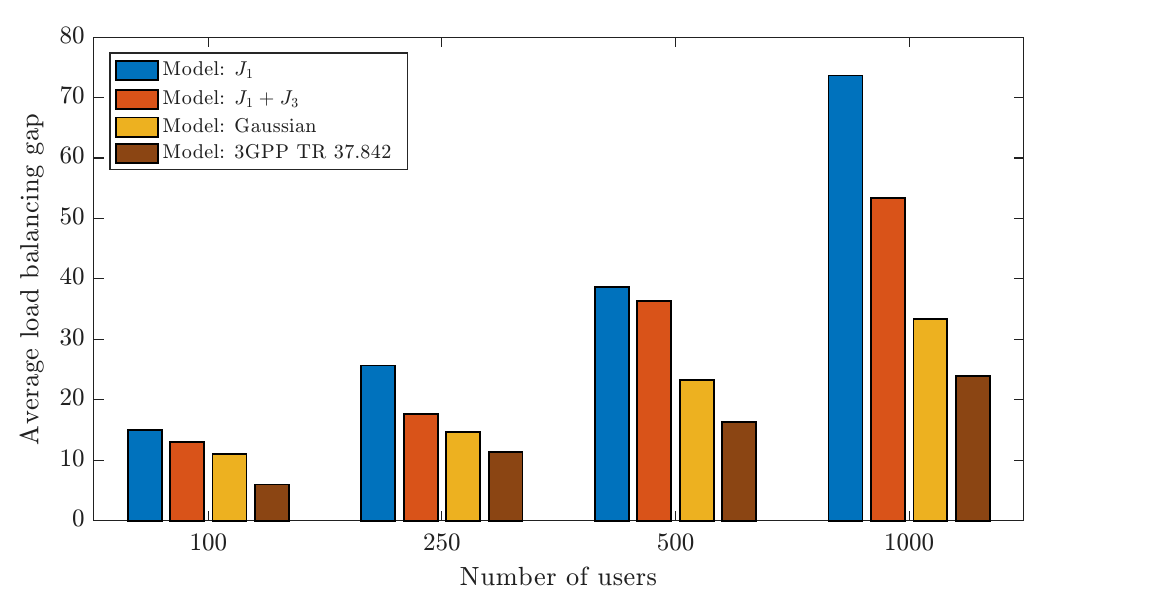}
    \caption{Average load balancing gap versus the number of users.}
    \label{fig:load_balance_comparison}
\end{figure}

Fig.~\ref{fig:load_balance_comparison} shows the average load balancing gap, defined as the difference between the maximum and minimum numbers of users served by active beams. The results indicate that the gap generally increases with the number of users. The 3GPP-TR 37.842 model achieves a smaller gap because it activates more beams, allowing users to be distributed more evenly. In contrast, the Bessel-based models use fewer beams, which leads to larger user groups and higher load imbalance. By comparing Fig.~\ref{fig:nab_comparison} and Fig.~\ref{fig:load_balance_comparison}, a clear trade-off can be observed between beam resource efficiency and load fairness. The combined $J_1+J_3$ model minimizes the number of active beams but may cause higher load imbalance, especially in dense scenarios. Conversely, the 3GPP-TR 37.842 model improves load balancing at the cost of activating more beams.

\subsection{SCGNR Distribution}

\begin{figure}[t]
    \centering
    \includegraphics[trim={1.7cm 8.8cm 0.0cm 9.0cm},clip,width=0.5525\textwidth]{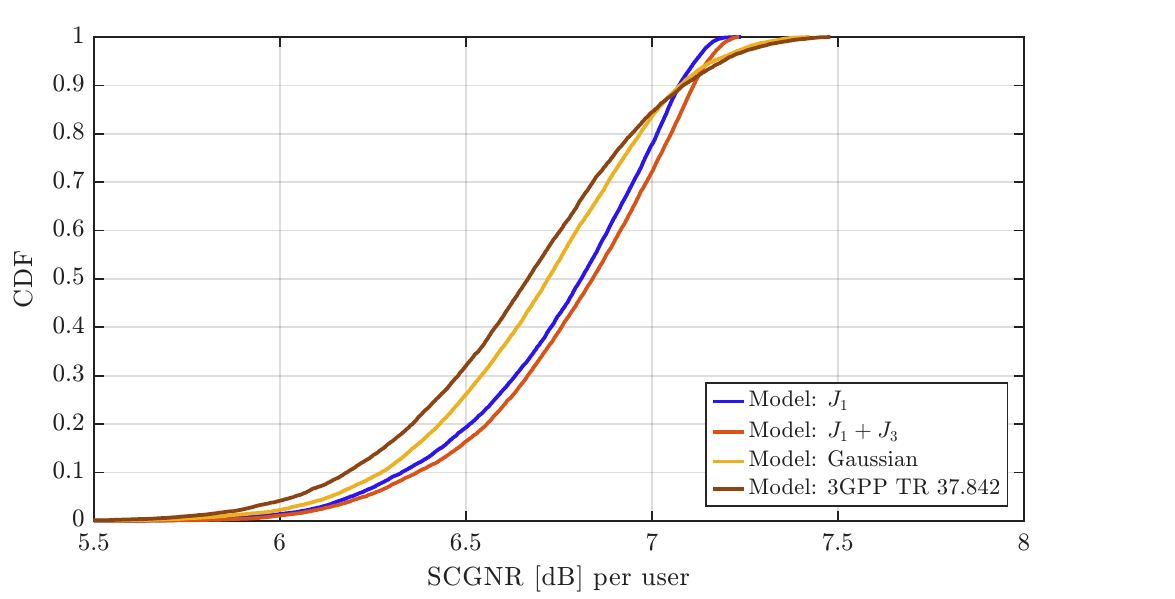}
    \caption{CDF of SCGNR for different antenna beam pattern models.}
    \label{fig:scgnr_cdf}
\end{figure}

Fig.~\ref{fig:scgnr_cdf} shows the CDF of the SCGNR for the considered antenna models. The SCGNR reflects the link-budget quality without explicitly considering inter-beam interference. A curve shifted to the right indicates better channel gain performance. The Bessel $J_1$ and combined $J_1+J_3$ models achieve better SCGNR distributions than the Gaussian and 3GPP-TR 37.842 models. This is because the Bessel-based patterns provide stronger gain around the main-lobe region for the obtained beam placement. In contrast, the Gaussian and 3GPP-TR 37.842 models show lower SCGNR values, especially in the lower-tail region of the CDF. This result confirms that the antenna gain model affects not only the beam grouping process but also the resulting channel gain experienced by users. Specifically, the wider main lobe of the Bessel-based models maintains higher antenna gains over a larger angular range before decaying sharply. This characteristic prevents cell-edge users from experiencing severe link budget degradation, thereby compressing the lower tail of the SCGNR CDF.

\subsection{SINR-Based Spectral Efficiency}

\begin{figure}[t]
    \centering
   \includegraphics[trim={0.15cm 7.4cm 0.6cm 7.9cm},clip,width=0.465\textwidth]{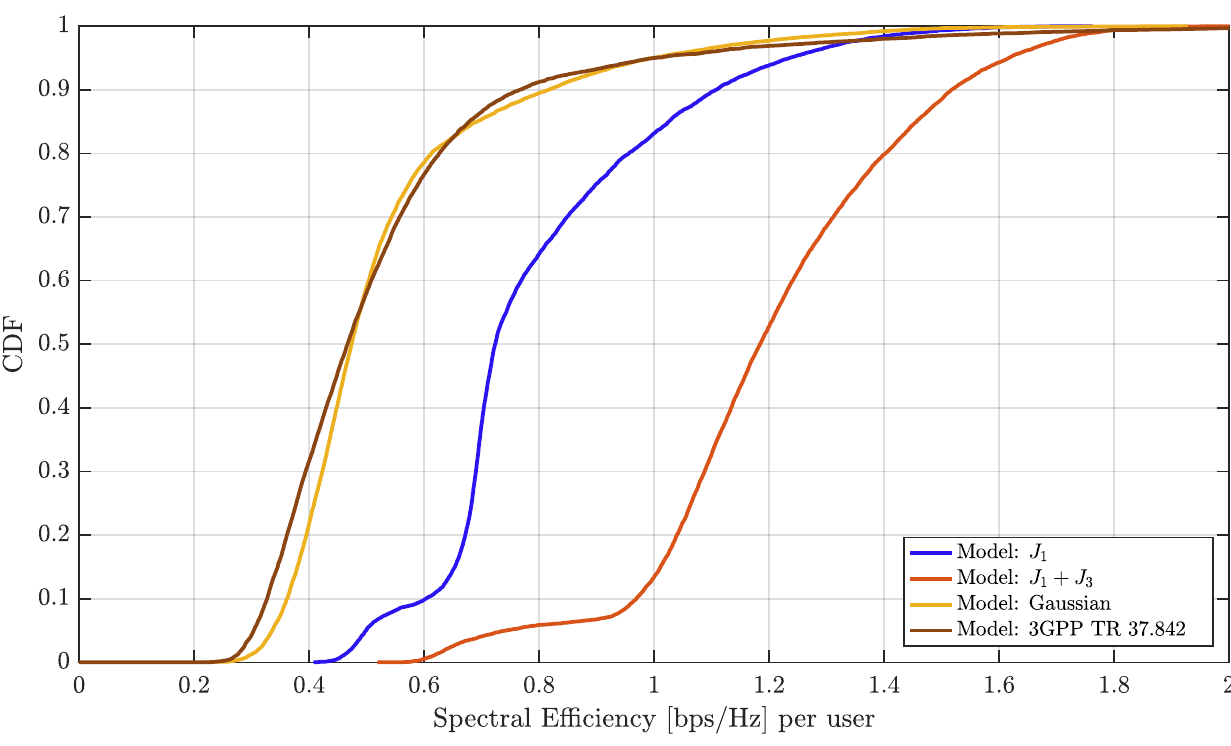}
    \caption{CDF of SINR-based spectral efficiency for different antenna models using Graph-based beam placement.}
    \label{fig:se_model_cdf}
\end{figure}

Fig.~\ref{fig:se_model_cdf} shows the CDF of the SINR-based spectral efficiency for different antenna models. Different from SCGNR, the spectral efficiency is computed from SINR and therefore captures the impact of inter-beam interference. This experiment is conducted to evaluate user-level spectral efficiency under different beam patterns. The results show that the combined $J_1+J_3$ model provides the best spectral efficiency distribution among the considered models. The Bessel $J_1$ model also achieves competitive performance, while the Gaussian and 3GPP-TR 37.842 models exhibit lower spectral efficiency. This confirms that a model with favorable main-lobe characteristics can improve the SINR-based rate performance. However, the difference between SCGNR and spectral efficiency also suggests that evaluating only the channel gain may not be sufficient. Overall, the $J_1+J_3$ model provides the most favorable performance in terms of active beam reduction and spectral efficiency, whereas the 3GPP-TR 37.842 model offers better load balancing due to its narrower effective coverage. 

\subsection{Impact of Number of Active Beams on Spectral Efficiency}

\begin{figure}[t]
    \centering
    \includegraphics[trim={1.7cm 8.8cm 0.0cm 9.0cm},clip,width=0.5525\textwidth]{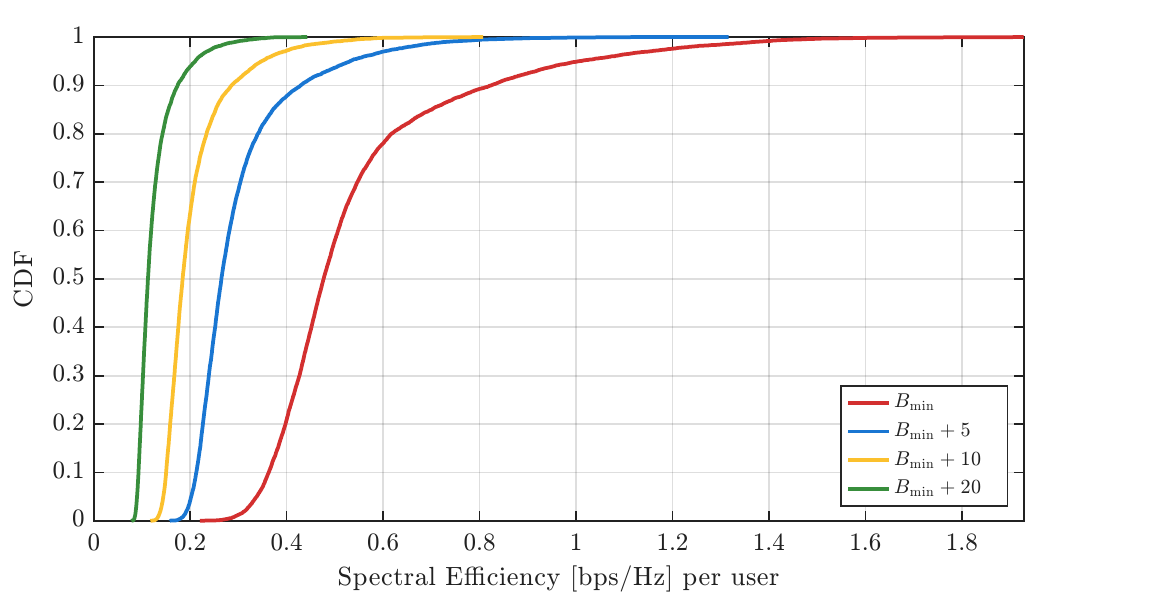}
    \caption{CDF of SINR-based spectral efficiency under different numbers of active beams using fixed-NBA K-means.}
    \label{fig:se_active_beam_cdf}
\end{figure}

To examine the effect of active-beam density, we compare the spectral efficiency under four configurations: $B_{\min}$, $B_{\min}+5$, $B_{\min}+10$, and $B_{\min}+20$, where $B_{\min}$ is obtained from the proposed graph-based placement under the considered antenna model. For each fixed number of beams, K-means clustering is applied to determine the beam groups. Fig.~\ref{fig:se_active_beam_cdf} shows the CDF of spectral efficiency for different active-beam numbers. The results indicate that activating more beams does not necessarily improve spectral efficiency. Although more beams can shorten the distance between users and their serving beam centers, they also divide the limited satellite transmit power and increase inter-beam interference under full frequency reuse. Under the considered full-frequency-reuse and equal-power setting, the $B_{\min}$ configuration achieves the best spectral efficiency. This confirms the importance of carefully controlling the number of active beams in LEO multibeam systems.

\section{Conclusion}
\label{sec:conclusion}

In this paper, we investigated the impact of antenna beam pattern models on fast beam placement in LEO satellite networks. Four representative antenna models were incorporated into a graph-based placement framework through their corresponding half-power angles. The beam placement problem was formulated as a Minimum Clique Cover problem and solved using graph coloring. Numerical results showed that antenna beam patterns significantly affect the number of active beams, load balancing, SCGNR, and SINR-based spectral efficiency. Wider beam patterns can reduce the required number of active beams but may increase load imbalance, while activating more beams does not necessarily improve spectral efficiency due to power sharing and inter-beam interference. These results highlight the importance of antenna-pattern-aware beam placement in LEO multibeam systems. Future work will consider multi-satellite scenarios and joint optimization of beam placement, beamwidth, frequency reuse, and power allocation.

\section*{Acknowledgment}
This research is funded by the Vietnam Ministry of Education and Training under project number CT2025.EA.BKA.20.

\bibliographystyle{IEEEtran}
\bibliography{IEEE}

\end{document}